\documentstyle{article}
\input epsf
\begin{document}
\def\beq{\begin{equation}}
\def\beql#1{\begin{equation}\label{#1}}
\def\eeq{\end{equation}}
\def\r{\rho}\def\ro{\hat\r}
\def\s{\sigma}\def\so{\hat\s}
\def\sv{\vec\s}\def\svo{\hat{\sv}}
\def\Pio{\hat\Pi}
\def\Io{\hat I}
\def\wv{\vec w}
\def\ket#1{\vert#1\rangle}
\def\bra#1{\langle#1\vert}
\def\qexpt#1{\langle#1\rangle_t}
\def\tr{\mbox{tr}}
\parbox{13 cm}
{
\begin{flushleft}
\vspace* {1.2 cm}
{\Large\bf
{
Single Qubit Estimation from Repeated Unsharp Measurements
}
}\\
\vskip 1truecm
{\large\bf
{
Lajos Di\'osi$^1$)
}
}\\
\vskip 5truemm
{
$^1$) Research Institute for Particle and Nuclear Physics\\
      H-1525 Budapest 114, POB 49, Hungary
}
\end{flushleft}
}
\vskip 0.5truecm
{\bf Abstract:\\}
{
\noindent
When estimating an unknown single pure qubit state, the optimum fidelity
is $2/3$. As it is well known, the value $2/3$ can be achieved in one
step, by a single ideal measurement of the polarization along a
random direction. I analyze the opposite strategy which is the
long sequence of unsharp polarization measurements. The evolution of
the qubit under the influence of repeated measurements is quite
complicated in the general case. Fortunately, in a certain
limit of very unsharp measurements the qubit will obey simple
stochastic evolution equations known for long under the name of
time-continuous measurement theory. I discuss how the outcomes of the
very unsharp measurements will asymptotically contribute to our
knowledge of the original qubit. It is reassuring that the fidelity
will achieve the optimum $2/3$ for long enough sequences of the unsharp
measurements.}


\section{Introduction}

Quantum measurement ${\bf M}$ means the procedure to obtain the value $\s$ of 
some hermitian observable $\so$ of the given quantum system. The {\it apriori}
state $\ro$ of the system transforms into the {\it aposteriori} state 
$\ro(\s)$ conditioned on the measurement outcome $\s$. The theory of
quantum measurement is well-known for projective (sharp) as well as for
non-projective (unsharp) measurements. There is, however, a further task 
beyond quantum measurement. One can consider the {\it apriori} quantum state 
$\ro$ of the given system as an additional object of inference 
\cite{Hel76,Hol82}. The estimation ${\bf E}$ is based on the measurement 
outcome $\s$. Hence the {\it estimate} state $\ro'(\s)$ becomes, similarly 
to the aposteriori state $\ro(\s)$, the function of the measurement outcome. 
This function depends on the estimation strategy ${\bf E}$ \cite{foo1}. 
The flowchart of quantum inference, consisting of the quantum measurement 
${\bf M}$ and of the estimation ${\bf E}$, is displayed on Fig. \ref{ME}. 
\begin{figure}[htbp]
\begin{center}
\epsfxsize=3in
\epsfbox{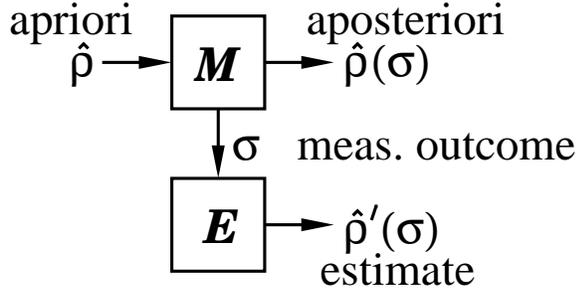}
\caption{Quantum inference: measurement $(M)$ and estimation $(E)$. 
\label{ME}}
\end{center}
\end{figure}
Unlike the theory of measurement, the theory of estimation has not so far
achieved a complete understanding. Most results are restricted for pure
apriori states. A completely unknown state $\ro$ can not be 
inferred from a {\it single} system: the {\it fidelity} of the estimate $\ro'$
will be poor.  If the apriori state $\ro$ is pure then the estimate
$\ro'$ must also be pure, and the simple bilinear expression 
$F=\tr[\ro'\ro]$ defines its fidelity. If we assume that the apriori 
pure $\ro$ is completely random then lower and upper limits become 
analytically calculable for the average fidelity $\bar F$ \cite{BM99}.
For a single two-state system (qubit) one obtains:
\beql{Flims}
\frac{1}{2} \leq \bar F \leq \frac{2}{3}~~.
\eeq
Any deliberate trial $\ro'$, when completely unrelated to $\ro$, will
yield the same worst value $1/2$. The best value can be attained in many 
ways. Let us, for instance, measure the Pauli-polarization matrix 
$\so$ along a single randomly chosen spatial direction. Let 
$\s=\pm1$ be the results of the {\it projective} measurement. 
It is then natural to identify the estimate pure state $\ro'(\s)$ with 
the standard {\it aposteriori} pure state $\ro(\s)$ taught in textbooks:
\beql{ropr}
\ro'(\s)=\ro(\s)\equiv\frac{\Io+\s\svo}{2}~~.
\eeq
The average fidelity over random apriori pure states $\ro$ is $2/3$.
No quantum measurement however involved  could 
improve on ${\bar F}=2/3$. It would make no sense to perform
a second projective measurement on the given single qubit.
We can, however, consider {\it non-projective} (unsharp) measurements 
\cite{Kra83,NC00} from the 
beginning. It makes sense to combine 
successive non-projective measurements on a single system \cite{AKS01} 
in order to improve fidelity. This I call sequential inference. Its
flowchart is shown on Fig. \ref{MEseq}.  
\begin{figure}[htbp]
\begin{center}
\epsfxsize=4in
\epsfbox{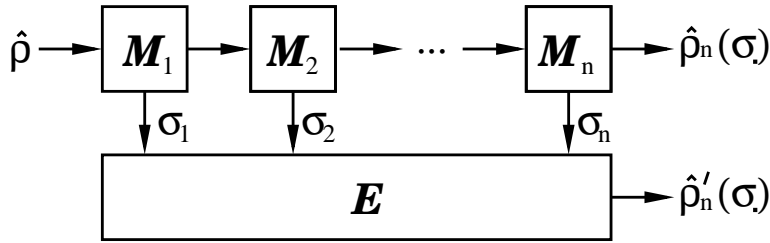}
\caption{Sequential quantum measurement and estimation.
\label{MEseq}}
\end{center}
\end{figure}
The question, discussed first in Ref.~\cite{Dio02}, is this. 
{\it For an unknown qubit $\ro=\ro^2$, do many $(n\!\!\gg\!\!1)$ unsharp 
measurements (of precision $\Delta\!\!\gg\!\!1$) of random polarizations 
$\so_1,\so_2\dots,\so_n$ allow an optimum estimate $\ro'$ (of fidelity 
$2/3$)?} We shall see that they do!

There is a particular limit of sequential inference which is tractable by
stochastic differential equations. This may be called continuous
inference (Fig. \ref{MEcont}). 
\begin{figure}[htbp]
\begin{center}
\epsfxsize=4in
\epsfbox{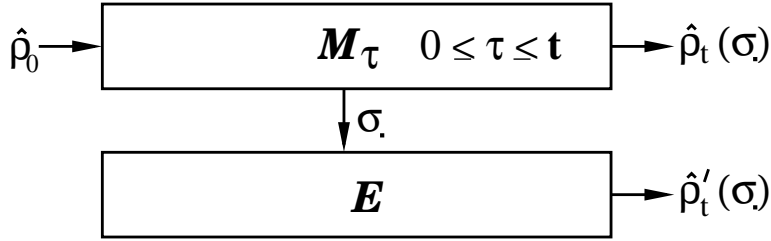}
\caption{Continuous quantum measurement and estimation.
\label{MEcont}}
\end{center}
\end{figure}
It is valid for very long sequences of very unsharp measurements. The elegant
differential equations of the continuous measurement have been known for 
long \cite{Dio88}. The more complex differential equations of the 
continuous estimation give little assistance, at least in their first derived
form \cite{Dio02}. To overcome their difficulty, I applied a trick. One
can consider a hypothetical apriori state whose hypothetical aposteriori
states are identical to the true estimate states. Hence the differential
equations of continuous measurement can be used to derive certain
properties like, in particular, the fidelity of the estimate state.  

\section{Estimation from POVM}

We approximate the exact eigenstates of a given hermitian 
observable $\so$ by approximate Gaussian projectors of precision $\Delta$: 
\beql{Pio}
\Pio(\s)=\frac{1}{\sqrt{2\pi\Delta^2}}
      \exp\left[-\frac{(\so-\s)^2}{2\Delta^2}\right]~~.
\eeq
They satisfy the completeness condition
\beql{compl}
\int\Pio(\s)d\s=\Io~~,
\eeq
and form a POVM \cite{Kra83,NC00}. The 
corresponding (non-projective) measurement of $\so$ will transform the 
apriori state $\ro$ into the following aposteriori state:
\beql{apost}
{\bf M}:~~\ro\longrightarrow\ro(\s)=\frac{\Pio^{1/2}(\s)\ro\Pio^{1/2}(\s)}
                              {\tr\left[\Pio(\s)\ro\right]}~~,
\eeq
where $\s$ is the random outcome of the measurement. It may take any
real value with the normalized probability density
\beql{p}
p(\s)=\tr\left[\Pio(\s)\ro\right]~~.
\eeq

The theory of (non-projective) measurements does not imply a theory 
for the estimate $\ro'$. One could mistakenly think the aposteriori state 
$\ro(\s)$ a reasonable estimate for the apriori state $\ro$. 
Unfortunately, the experimenter has no access to it. He/she infers 
the measured value $\s$ and it is, contrary to the projective 
measurement (\ref{ropr}), not enough to derive the aposteriori state. 
It is only sufficient to identify the approximate projector $\Pio(\s)$. 
Its normalized form can be a reasonable estimate:
\beql{est}
\ro'(\s)=\frac{\Pio(\s)}{\tr\Pio(\s)}~~.
\eeq
This is a mixed state. If the apriori states $\ro$ are unknown
pure states then the estimate should also be pure. To this end, the 
experimenter must refine his/her first choice (\ref{est}). The   
estimate will be one of the pure eigenstates of the mixed state 
estimate (\ref{est}), chosen randomly with probability equal to the 
corresponding eigenvalue. (The optimum estimate would be the most 
probable eigenstate \cite{Ban01}.) The bilinearity of
fidelity $\tr[\ro'\ro]$, valid originally between two pure states, 
will be preserved for the expected fidelity of our estimates: 
\beql{F}
F=\int\tr\left[\ro'(\s)\ro\right]p(\s)d\s
 \equiv \mbox{E }\tr[\ro'(\s)\ro]~~,
\eeq
where $\ro'$ is defined by (\ref{est}) and $\mbox{E}$ stands for 
stochastic expectation value. 

\section{A useful trick}

The expected fidelity (\ref{F}) of our estimate (\ref{est}) has been 
expressed in terms of the apriori $\ro$ and the estimate state $\ro'(\s)$. 
There is an alternative expression depending on a {\it hypothetic} 
aposteriori state $\ro^?(\s)$ and on the true apriori state $\ro$. 
The trick is that the true estimate state $\ro'(\s)$ can be identified
by the hypothetic aposteriori state:
\beql{trick}
\ro'(\s)=\ro^?(\s)~~,
\eeq
where the
state $\ro^?(\s)$ results from a hypothetic measurement of the POVM
$\Pio(\s)$ on a completely mixed (hypothetic) apriori state 
$\ro^?=\Io/2$. Indeed, the measurement (\ref{apost}) yields
\beql{apost?}
{\bf M}:~~\ro^?=\frac{\Io}{2}\longrightarrow\ro^?(\s)=\frac{\Pio(\s)}
                              {\tr\Pio(\s)}~~,     
\eeq
while
\beql{p?}
p^?(\s)=\tr\left[\Pio(\s)\ro^?\right]=\frac{1}{2}\tr\Pio(\s)
\eeq
is the probability distribution of the outcome. In the expression
(\ref{F}) of fidelity we can thus replace $\ro'(\s)$ by $\ro^?(\s)$ and
$p(\s)$ by $2p^?(\s)\tr[\ro^?(\s)\ro]$ yielding:
\beql{F?}    
F=2\int\left(\tr\left[\ro^?(\s)\ro\right]\right)^2 p^?(\s)d\s
\equiv 2\mbox{E }\!\!\left(\tr\left[\ro^?(\s)\ro\right]\right)^2~~.
\eeq
Note that the stochastic average is to be taken with the hypothetical
probability distribution $p^?(\s)$ instead of the true $p(\s)$.
The new expression (\ref{F?}) contains the (hypothetical) aposteriori 
state while the old formula (\ref{F}) contained the (true) estimate state. 
Finally, we have to average the fidelity (\ref{F?}) over random pure qubit 
states $\ro$:
\beql{Fav}
\bar F=\frac{1}{3}+\frac{1}{3}\mbox{E }\tr[\ro^?(\s)]^2~~.
\eeq
This formula of the average fidelity depends completely on the 
{\it purity} $\tr[\ro^?(\s)]^2$ of the hypothetic
aposteriori state $\ro^?(\s)$. Purity's minimum value is $1/2$ for the 
totally mixed state while its maximum is $1$ for a pure state.

\section{Sequential inference}

According to Fig. \ref{MEseq}, we apply $n$ unsharp measurements
$\Pio_1(\s_1),\dots,\Pio_n(\s_n)$ (\ref{Pio}-\ref{p}) of the respective
observables $\so_1,\dots,\so_n$ which are polarizations along independent
random directions. Using the shorthand notation $(\s_1,\dots,\s_n)=(\s_.)$ 
for the measurement outcomes, the aposteriori and the estimate states will 
be denoted by $\ro_n(\s_.)$ and $\ro'_n(\s_.)$, respectively.  To estimate 
the state from the measurement outcomes, we can follow the recipe of single 
inference. The above sequence of $n$ measurements constitute a single 
(complicated) measurement. It has its POVM $\Pio_n(\s_.)$ \cite{Dio02}. 
Extending the estimation strategy from single measurement, we introduce the 
mixed state estimate
\beql{est_n}
\ro'_n(\s_.)=\frac{\Pio_n(\s_.)}{\tr\Pio_n(\s_.)}
\eeq
whose eigenstates, like in case of (\ref{est}), will be the pure state 
estimates. Same considerations that led to fidelities 
(\ref{F},\ref{Fav}) apply invariably.
We can, for instance, write the average fidelity in terms of the 
aposteriori state $\ro^?_n(\s_.)$ emerging from a hypothetical
apriori qubit state $\ro^?\equiv\ro^?_0=\Io/2$:
\beql{Fav_n}
\bar F_n=\frac{1}{3}+\frac{1}{3}\mbox{E }\tr[\ro_n^?(\s_.)]^2~~.
\eeq
It is obvious that $\bar F_0=1/2$, and we expect $\bar F_n$ is a 
monotone function of $n$. We shall prove that $\ro_n^?(\s_.)$ tends to be
pure for large $n$ hence $\bar F_n$ attends the optimum (\ref{F}).

\section{Continuous inference}

We assume long sequences of very unsharp measurements:
\beql{lim}
n\gg1~,~~~\Delta\gg1~~.
\eeq
The asymptotic limit \cite{Bar86,Dio88}
\beql{cont}
n,\Delta\longrightarrow\infty~,~~~~\frac{n}{\Delta^2}=\mbox{const}
\eeq
will be called the `continuum limit'.
Formally, let us count the succession of 
measurements as if they happened at constant rate $\nu=12/\Delta^2$.
Accordingly, we replace the discrete parameter $n$ by the 
continuous time:
\beql{t}
t=\frac{12n}{\Delta^2}~~.
\eeq
We consider all quantities as continuous functions of $t$,  
coarse-grained on scales $\gg\!\!1/\nu$ involving many
measurements. In this limit an approximate theory emerges
in the form of markovian stochastic differential equations. 
(The theory becomes exact in the continuum limit.)
The aposteriori state 
satisfies the conditional (or selective) master equation:
\beql{dro_t}
\frac{d\ro_t}{dt}=-\frac{1}{2}[\svo,[\svo,\ro_t]~]
                  +\{\svo-\qexpt{\svo},\ro_t\}\wv_t~~,
\eeq
where $\qexpt{\svo}=\tr[\svo\ro_t]$.
We have suppressed denoting the functional dependence of $\ro_t$
on the outcomes $\{\s_\tau; 0\leq\tau\leq t\}$. 
The $\wv_t$ is the standard 
isotropic white-noise and the equation must be interpreted in the 
sense of the Ito stochastic calculus.
There is a second stochastic differential equation for the outcome: 
\beql{sv_t}
\sv_t=\qexpt{\svo} + \frac{1}{2}\wv_t~~.
\eeq
The features of the above equations have been well understood \cite{foo2}. 
This is not yet achieved for the differential equations, coupled to
(\ref{dro_t},\ref{sv_t}) via the noise $w_t$, which govern the estimate 
$\ro_t'$ \cite{Dio02}. Coming back to the solution $\ro_t$, it is known that
for all initial states $\ro_0$, including mixed ones, the aposteriori state
$\ro_t$ becomes asymptotically pure for long times \cite{Kor00,DTPW99}. 
This assures the saturation of average fidelity (\ref{Fav_n}), as proven in 
the next section. 

\section{Saturation of fidelity}  

We calculate the average fidelity $\bar F_t$. It corresponds to the 
(coarse-grained) $n-$dependent fidelity $\bar F_n$ (\ref{Fav_n})
via $t=12n/\Delta^2$. The latter
requires the knowledge of the hypothetical aposteriori state  
$\ro^?_t$ evolving from the hypothetical initial apriori state 
$\ro=\ro^?_0=\Io/2$. The stochastic `master' equation (\ref{dro_t}) 
yields a certain diffusion process for the purity $\tr[\ro_t^?]^2$.
For long times it will approach the unity, therefore the aposteriori state 
$\ro_t^?$ becomes asymptotically pure. My Monte-Carlo calculations 
have shown that the purity $\tr[\ro_t^?]^2$ is dominated by the drift term. 
Ignoring diffusion, the analytic solution is possible \cite{Dio02}:
\beql{drift}
\tr[\ro_t^?]^2=\frac{1}{2}+\frac{1}{2}\frac{e^{8t}-1}{e^{8t}-1/3}~~.
\eeq  
Let us restore the original variable $n=t\Delta^2/12$
and substitute the above result into the expression (\ref{Fav_n}):
\beql{sat}
\bar F_n=\frac{1}{3}+\frac{1}{3}\mbox{E }\tr[\ro_t^?]^2
        =\frac{1}{2}+\frac{1}{6}\frac{e^{96n/\Delta^2}-1}
                                     {e^{96n/\Delta^2}-1/3}~~.
\eeq
The average fidelity approaches the optimum value $2/3$ after
a characteristic number $n\sim\Delta^2/96$ of unsharp measurements. 
Recalling the conditions (\ref{lim}) we conclude that our result
is valid for {\it very} unsharp measurements, i.e.,
$\Delta$ must be at least $\sim\!\!\sqrt{96}$-times the
natural scale of polarization $\s$. 

\section{Concluding remarks}

I have proven that a very long sequence of very unsharp polarization
measurements on a single qubit will provide the optimum fidelity $2/3$
in estimating the unknown apriori (pure) state. The details, including the
strategy of estimation, are given in Ref.~\cite{Dio02}.
The result represents the first steps in extending the
theory of continuous quantum measurement to continuous quantum estimation
which altogether may constitute a future theory of continuous 
quantum inference. (For a related concept, restricted for Gaussian states,
see Ref.~\cite{DTPW99}.) This may be of interest every time one is 
accumulating and analyzing information from low rate quantum inference 
(e.g.: eavesdroppers of secret quantum communication, tomography with low 
detection efficiency \cite{VR89,Ban99}, cloner of $n\!\!\gg\!\!1$ identical 
qubits into $n+1$). One might be able to extend the concept of continuous 
estimation for pure states of non-trivial apriori distribution, see e.g.
the issues in Ref.~\cite{Ber02}. Whether it offers optimum fidelities we do 
not know for the moment.
\vskip .5truecm
This work was supported by the Hungarian OTKA Grant 32640. 




\begin{thebibliography}{1}
\bibitem{foo1} The limitations as well as the 
optimization of state determination are in the focus of recent investigations
\cite{MP95DBE98VLPT99,BM99,Ban01} especially in the field of quantum
information and communication \cite{NC00}.
\bibitem{foo2} The `conditional' master equation \cite{Gis84} and its 
coupling to the measured value \cite{Dio88} were known long ago (see also 
\cite{Bel89Pea89GP92Car93WM93}) as the ultimate formalism of earlier 
continuous measurement models \cite{Bar86,Men79GRW86CM87}.
\end{thebibliography}
\end{document}